\documentclass[prl,aps,twocolumn,superscriptaddress]{revtex4}
\usepackage{amsfonts}

\usepackage{dcolumn}
\usepackage{bm}
\usepackage{tikz}
\usepackage[colorlinks=true,citecolor=blue,urlcolor=blue]{hyperref}
\usepackage{amsmath,amsfonts,amssymb,times,natbib}
\usepackage[standard]{ntheorem}
\usepackage{threeparttable}

\begin{document}

\title{Experimental Simultaneous Learning of Multiple Non-Classical Correlations}

\author{Mu Yang}
\affiliation{CAS Key Laboratory of Quantum Information, University of Science and Technology of China, Hefei 230026, People's Republic of China}
\affiliation{CAS Center For Excellence in Quantum Information and Quantum Physics, University of Science and Technology of China, Hefei 230026, People's Republic of China}

\author{Chang-liang Ren}
\affiliation{Center for Nanofabrication and System Integration, Chongqing Institute of Green and Intelligent Technology, 400714, Chinese Academy of Sciences, People’s Republic of China}

\author{Yue-chi Ma}
\affiliation{ Center for Quantum Information, Institute for Interdisciplinary
	Information Sciences, Tsinghua University, Beijing 100084, People's Republic of China}
\affiliation{Shenzhen Institute for Quantum Science and Engineering and Department of Physics, Southern University of Science and Technology, Shenzhen 518055, China}

\author{Ya Xiao}
\affiliation{Department of Physics, Ocean University of China, Qingdao 266100, People's Republic of China}

\author{Xiang-Jun Ye}
\affiliation{CAS Key Laboratory of Quantum Information, University of Science and Technology of China, Hefei 230026, People's Republic of China}
\affiliation{CAS Center For Excellence in Quantum Information and Quantum Physics, University of Science and Technology of China, Hefei 230026, People's Republic of China}

\author{Lu-Lu Song}
\affiliation{Shenzhen Institute for Quantum Science and Engineering and Department of Physics, Southern University of Science and Technology, Shenzhen 518055, China}
\affiliation{Shenzhen Key Laboratory of Quantum Science and Engineering,
Southern University of Science and Technology, Shenzhen, 518055, China}

\author{Jin-Shi Xu}\email{jsxu@ustc.edu.cn}
\affiliation{CAS Key Laboratory of Quantum Information, University of Science and Technology of China, Hefei 230026, People's Republic of China}
\affiliation{CAS Center For Excellence in Quantum Information and Quantum Physics, University of Science and Technology of China, Hefei 230026, People's Republic of China}

\author{Man-Hong Yung}\email{yung@sustc.edu.cn}
\affiliation{Shenzhen Institute for Quantum Science and Engineering and Department of Physics, Southern University of Science and Technology, Shenzhen 518055, China}
\affiliation{Shenzhen Key Laboratory of Quantum Science and Engineering,
Southern University of Science and Technology, Shenzhen, 518055, China}
\affiliation{Central Research Institute, Huawei Technologies, Shenzhen, 518129, China}

\author{Chuan-Feng Li}\email{cfli@ustc.edu.cn}
\affiliation{CAS Key Laboratory of Quantum Information, University of Science and Technology of China, Hefei 230026, People's Republic of China}
\affiliation{CAS Center For Excellence in Quantum Information and Quantum Physics, University of Science and Technology of China, Hefei 230026, People's Republic of China}

\author{Guang-Can Guo}
\affiliation{CAS Key Laboratory of Quantum Information, University of Science and Technology of China, Hefei 230026, People's Republic of China}
\affiliation{CAS Center For Excellence in Quantum Information and Quantum Physics, University of Science and Technology of China, Hefei 230026, People's Republic of China}

\date{\today}
\begin{abstract}
{Non-classical correlations can be regarded as resources for quantum information processing. However, the classification problem of non-classical correlations for quantum states remains a challenge, even for finite-size systems. Although there exist a set of criteria for determining individual non-classical correlations, a unified framework that is capable of simultaneously classifying multiple correlations is still missing. In this work, we experimentally explored the possibility of applying machine-learning methods for simultaneously identifying non-classical correlations. Specifically, by using partial information, we applied artificial neural network, support vector machine, and decision tree for learning entanglement, quantum steering, and non-locality. Overall, we found that for a family of quantum states, all three approaches can achieve high accuracy for the classification problem. Moreover, the run time of the machine-learning methods to output the state label is experimentally found to be significantly less than that of state tomography.
}

\end{abstract}

\maketitle

\textit{Introduction}.---In 1935, Einstein, Podolsky, and Rosen (EPR)~\cite{Einstein} questioned the completeness of quantum mechanics (QM), as the theory seems to allow ``spooky action at a distance" (known as EPR paradox). In quantum information science, much efforts have been devoted to achieve a deeper understanding of EPR's paradox in terms of non-classical correlations, such as quantum entanglement~\cite{Horodecki}, EPR steering~\cite{Brunner}, and Bell non-locality \cite{Cavalcanti}. The various relationships among different non-classical correlations not only shape the foundation of the quantum theory, but also find themselves many interesting applications.

The question is, how may one characterize the non-classical correlation for any given quantum state? To tackle such a problem, there are several challenges. (i) Even though various mathematical criteria and inequalities constraining non-classical correlations are known, for general multipartite states, the classification of entanglement, EPR steering, or Bell non-locality are generally computationally-hard problems~\cite{Gurvits,Huang,Batle}. (ii) Many existing methods require the information of the whole density matrix; experimentally, a full quantum state tomography would be required. (iii) Each type of non-classical correlation has a different set of criteria; it is not known whether one can, simultaneously, classify all the non-classical correlations, based on the {\it same} set of measurements or observables.


On the other hand, machine learning (ML) represents a branch of artificial intelligence, aiming at producing a predictive function or a computer program based on a set of training data~\cite{Mitchell,Devroye,Vapnik,Friedman}. Beyond industrial applications, ML has created a profound impact on quantum information science, leading to a new research field, where many progresses have been achieved, including Hamiltonian learning~\cite{Wiebe}, automated quantum experiments search \cite{Krenn}, identification of phase transition~\cite{Schoenholz}, topological phase of matter~\cite{Zhang} etc.

Recently, binary classification of quantum correlations have been achieved using the tools of ML, such as determination of entanglement~\cite{Deng}, and separability~\cite{Lu,Ma,Jin}. Our goal here is to take one step further and explore if ML can be applied, in both theoretical and experimental setting, to {\it simultaneously} characterize multiple non-classical correlations with only partial information about the given quantum state.

\begin{figure}[t!]
	\begin{center}
		\includegraphics[width=0.98\columnwidth]{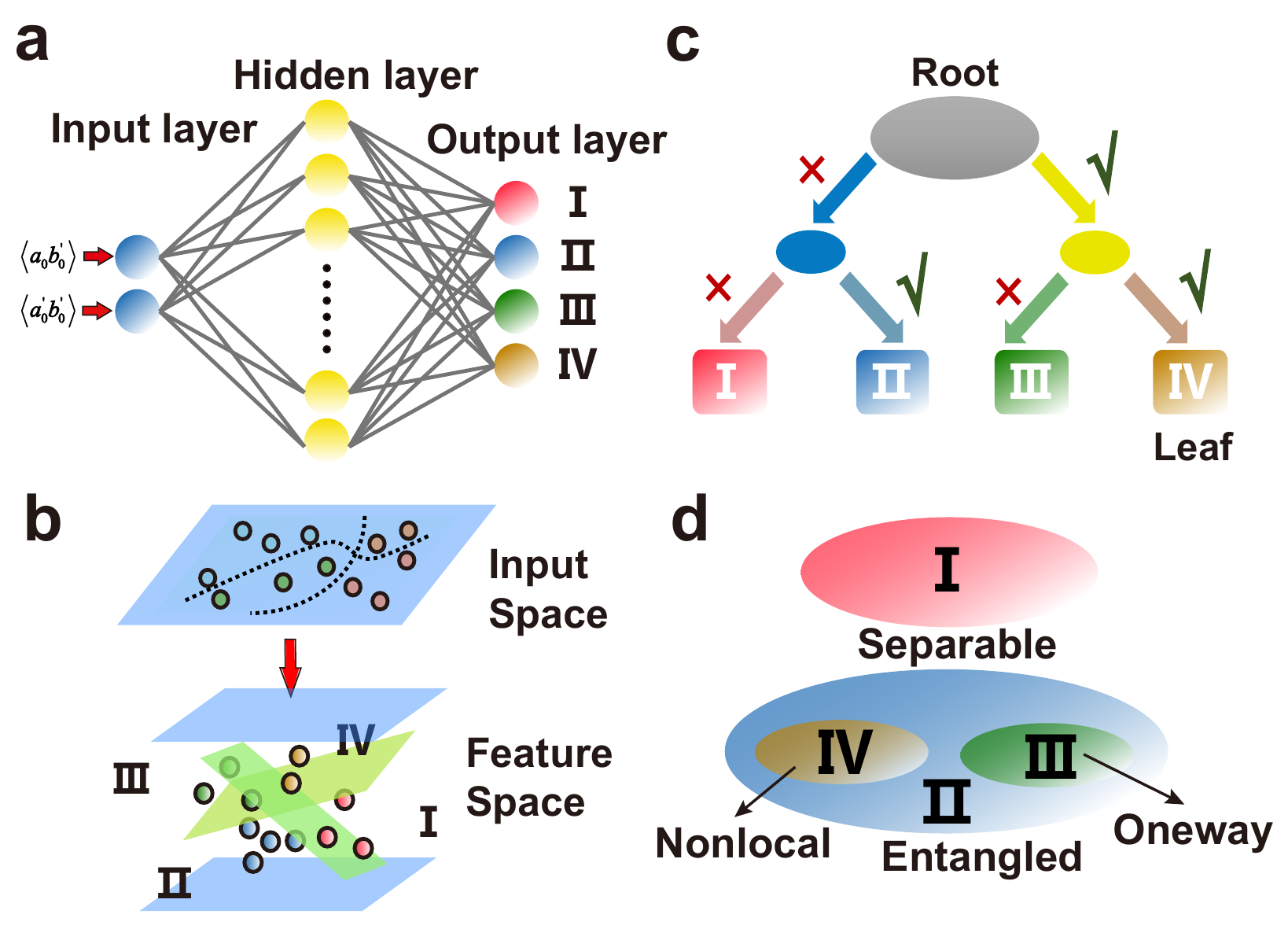}
		\caption{Quantum correlation is divided into 4 categories by three typical ML models. \textbf{a.}The structures of artificial neural network (ANN). \textbf{b.} The support vector machine (SVM). \textbf{c.} The decision tree (DT). \textbf{d.} The Venn diagram of these four categories, in which I, II, III and IV represent the separable state, entangled state, one-way steerable state and Bell nonlocal state.}		
		\label{MLM}
	\end{center}
\end{figure}

In addition to binary classifiers, we experimentally constructed a {\it statistical unified witnesses} for simultaneous characterizing different classes of multiple non-classical correlations through machine learning.
Specifically, we compared three different multi-label state classifiers using three different ML methods (see FIG.~\ref{MLM}), including artificial neural network (ANN), support vector machine (SVM) and decision tree (DT), where each classifier only takes partial information for each member in a family of quantum states. The label on the state is determined by using the positive partial transpose (PPT) criterion~\cite{PPT}, steering radius in two settings~\cite{sun2016experimental} and Clauser-Horne-Shimony-Holt (CHSH) inequalities~\cite{CHSH}, which are applied only to the training set but not the testing set.

\textit{Labeling the training states}.---For the purpose of demonstration, we shall focus on a family of quantum states, for which we can label unambiguously the class of non-classical correlations for each member. For feature extraction, we use partial information (two observables) of the quantum states, instead of the whole density matrix,
 \begin{equation}\label{ab}
\langle{a}_{0}{b}^{\prime}_{0}\rangle,\langle{a}^{\prime}_{0}{b}_{0}\rangle,
\end{equation}
to be the input, where ${a}_{0}={\sigma}_{z}$, ${a}^{\prime}_{0}={\sigma}_{x}$, ${b}_{0}=({\sigma}_{z}-{\sigma}_{x})/\sqrt{2}$, ${b}^{\prime}_{0}=({\sigma}_{z}+{\sigma}_{x})/\sqrt{2}$.

\begin{figure*}[t]
	\begin{center}
		\includegraphics[width=1.85 \columnwidth]{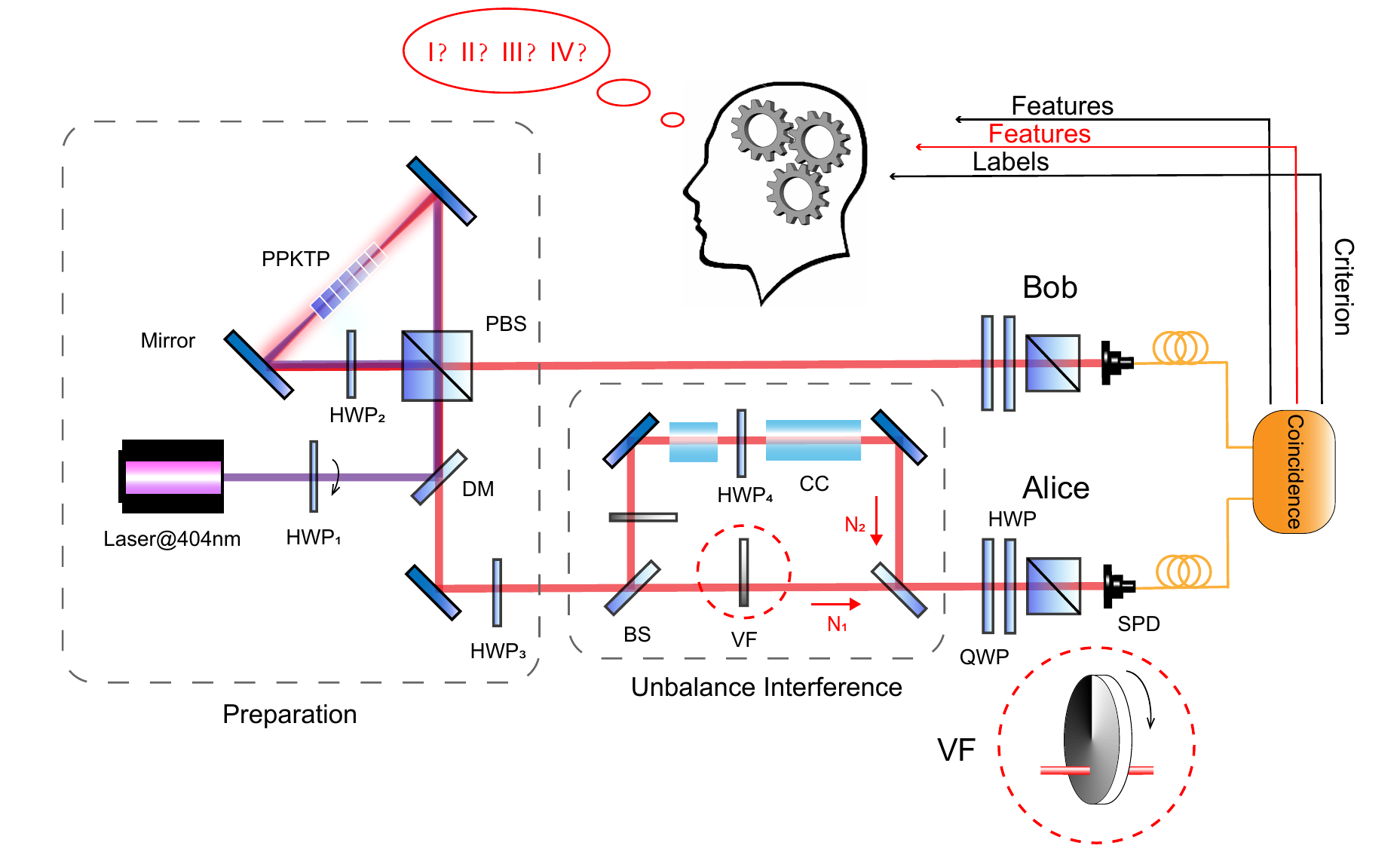}
		\caption{\textbf{Experimental setup}. A pair of polarization-entangled photons are generated in the preparation stage by pumping a type-II PPKTP crystal located in a Sagnac interferometer with an ultraviolet laser at 404 nm. The polarization state of the pump light is rotated by a half-wave plate (HWP$_{1}$). The dual-wavelength HWP$_{2}$ set to be $45^{\circ}$ in the Sagnac interferometer is used to exchange the polarizations of the pump light and down-converted photons. The HWP$_{3}$ set to be $45^{\circ}$ is used to change the form of the entangled state. One of the photons passes through an unbalanced interferometer and is sent to Alice. HWP$_{4}$ is set at $ 22.5^{\circ} $ and two sufficiently long calcite crystals (CCs) with the length of the last one being two times larger than the first one to completely destroy the coherence in different components. Two variable filters (VF, in red dotted line circle) are used to control the relative amplitudes between these two arms. The other photon is sent directly to Bob. Quarter-wave plates (QWPs), HWPs and polarization beam splitters (PBSs) on both sides of Alice and Bob are used for observables and tomography measurement. The photons are detected by single-photon detectors (SPDs), and the signals are sent for coincidence. The training and testing procedures are marked with black and red colors respectively. In the training case, criteria of quantum correlations are determined from tomographic data and the results are sent to the ML models together with the two characteristic features obtained from partial information measurement. In the testing case, only the two measured characteristic features are used to predict the labels.}
		\label{setup}
	\end{center}
\end{figure*}

Specifically, the family of quantum states under investigation is of the following form, which has been applied for demonstrating one-way steering~\cite{oneway}:
\begin{equation}\label{states}
\rho_{AB}(p,\theta)=p \vert \psi_\theta \rangle\langle \psi_\theta \vert+(1-p)I_{A}/2\otimes\rho_{B}^{\theta},
\end{equation}
where $\vert\psi_\theta\rangle =\cos\theta\vert 00\rangle+\sin\theta\vert 11\rangle$, and $ \rho_{B}^{\theta} = {\rm tr}_{A} (\vert \psi_\theta \rangle\langle \psi_\theta\vert)$. The members in the two-parameter family are characterized by the combination of the two parameters $\theta \in (0,2\pi)$, $p \in (0,1)$. 

The non-classical correlations for the testing state $\rho_{AB}(p,\theta)$ can be determined by the following rules: (i) for separability, e.g. through PPT criterion, one can show that whenever $p<1/3$, the states are separable. Otherwise, they are entangled. (ii) for one-way steering, within the range, $1/\sqrt{2}<p<1/\sqrt{1+\sin^{2}2\theta}$, the states are one-way steerable~\cite{J.Bowles}. (iii) for non-locality, whenever $p>1/\sqrt{1+\sin^{2}2\theta}$, the states become non-local~\cite{su2016beating}. Each set of the features are associated with a label in the set {I, II, III, IV}, shown as FIG. \ref{MLM}d.






\textit{Experimental setting}.---The experimental setup is shown in FIG. \ref{setup}. Photon pairs entangled in the polarization basis $\{ H,V\}$ ($H$ and $V$ represent horizontal and vertical polarization, respectively) are created through a type-II spontaneous parametric down conversion in a 20 mm-long periodical {KTiOPO$_{4}$} (PPKTP) crystal, which is located on the Sagnac interferometer~\cite{fedrizzi2007wavelength} and pumped by a 404 nm continuous-wave diode laser. In order to generate the entangled state, $\cos\theta\vert HH\rangle+\sin\theta\vert VV\rangle$, a half-wave plate (HWP$_1$) is used to control the parameter $\theta$ by rotating the polarization of pump laser. One of the two photons is sent to an unbalanced interferometer (UI) while the other photon is sent to Bob directly. 

In the UI, the photon is separated into two paths by a beam splitter (BS). The state in one of the paths remains unchanged. Two sufficiently long calcite crystals (CCs) with HWP$_4$ set to be $22.5^{\circ}$ between them are placed on the other path to completely destroy the coherence between different components. Two variable filters (VFs) are used to manipulate the relative photon counts ($N_1$, $N_2$) between these two paths, and the parameter $p$ can be demonstrated as $p=N_{1}/(N_{1}+N_{2})$. Combining these two paths into one, arbitrary two-qubit states $ \rho_{AB}(p,\theta) $ can be prepared. 

A quarter-wave plate (QWP), a HWP, and a polarization beam splitters (PBS) on both sides are used for quantum state tomography (16 projective measurements) and two characteristic observables (features, partial information) measurements (8 projective measurements). The integration time for each measurement is 5~s, and the maximal counts are about 60350. The labels can be deduced from the reconstructed density matrix $\rho_{AB} (p, \theta)$, by using PPT criteria, steering radius in two settings (see Supplementary Material (SM)~\cite{supplementry} for details) and CHSH inequalities. We implemented two independent experiments with different settings ($\theta_{2th}\to\theta_{1th}+\delta\theta$) to obtain a training set and a test set respectively, which are used to train and test the ML models.

\begin{figure}[t]
	\begin{center}
		\includegraphics[width=0.95\columnwidth]{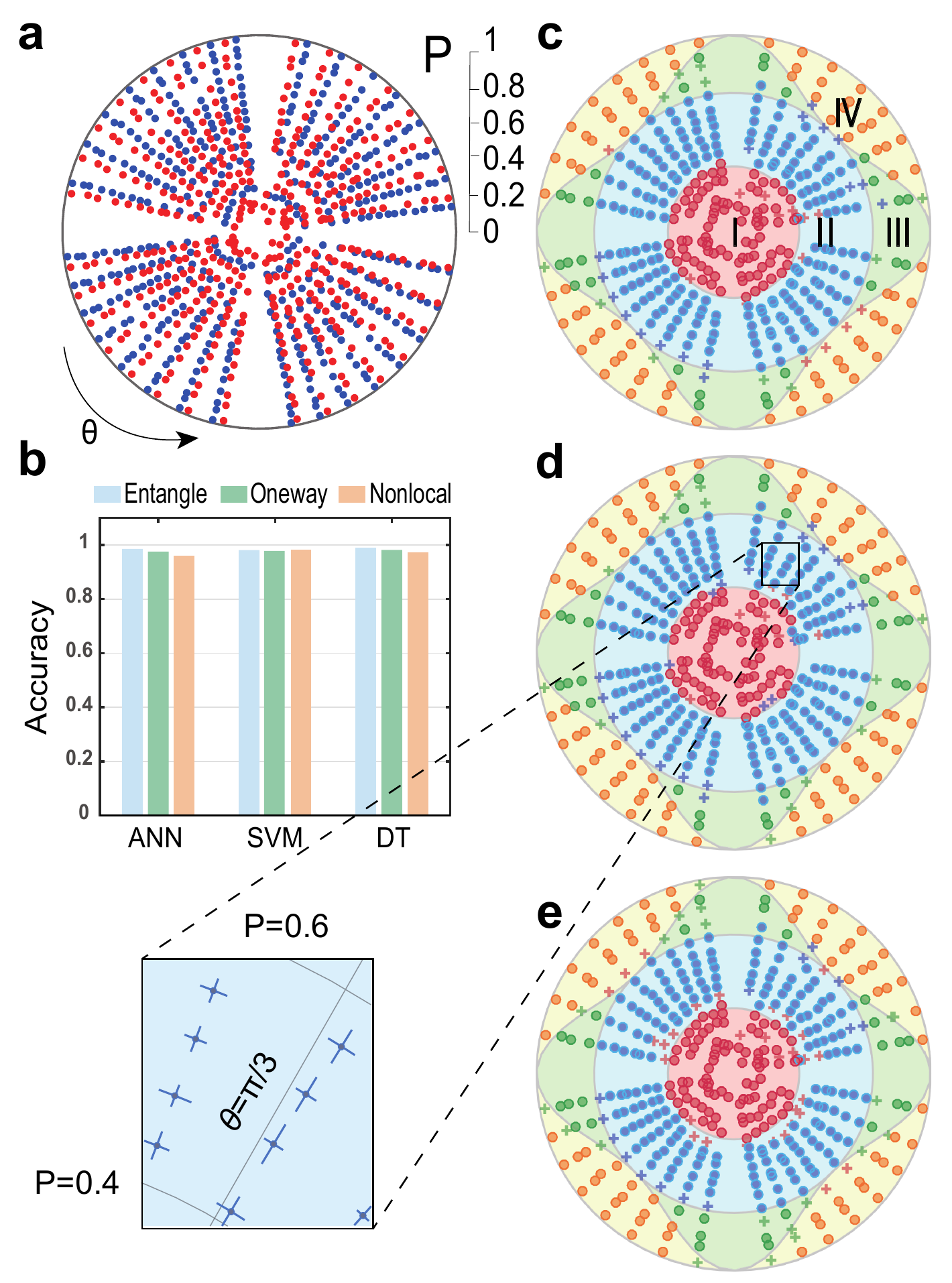}
		\caption{
		    {\bf a.} The training data (blue dots) and test data (red dots).
			{\bf b.} The accuracies of ANN, SVM and DT, when distinguish whether a state is entangled, one-way steerable, or Bell nonlocal.			
			The test results for multi-label classification with ANN, SVM and DT are shown in {\bf c.}, {\bf d.} and {\bf e.} with the accuracies of $90.11\%$, $90.11\%$ and $85.17\%$, respectively. The parameter $\theta$ varies from $0$ to $2\pi$ while the visibility $p$ changes from $0$ to $1$. The space is divided into four parts by the gray theoretical boundary, in which states in I, II, III and IV are theoretical predicted to be separable, entangled, one-way steerable and Bell nonlocal, respectively. ``$\times$" represent the mistakenly labeled states via ML models compared with those labeled by traditional criteria. The error bars are smaller than the size of the dots. The inset shows the enlargement of the corresponding region in the black pane. The error bars are due to the Poissonian counting statistics.
		}	
		
		\label{distribution}
	\end{center}
\end{figure}

The fidelities (${ Tr}[\sqrt{\sqrt{\rho_{AB}(p,\theta)}~\rho_{ \rm exp}\sqrt{\rho_{AB}(p,\theta)}}]^2$) between the theoretical physical states $\rho_{AB}(p,\theta)$ and experimental states $\rho_{\rm exp}$ are higher than $99\%$, so that the reconstruction error of the matrix does not affect the labels of the states determined by the traditional criteria. To train the input data, we prepared 445 states as the training set (shown as blue dots in FIG. \ref{distribution}a in the $(p, \theta)$ space). 
For the trained models, we experimentally prepared a different set of 455 states for testing, which are shown as red dots in FIG. \ref{distribution}a. The $(p, \theta)$ are calculated through minimizing the value of $Tr[\rho_{AB}(p, \theta)-\rho_{\rm exp}]$.

\textit{Experimental results}.---The two characteristic features of the testing states are then sent to different ML models. First, we consider applying the three ML models to perform {\it binary} classification (YES/NO labels). We found that the accuracy can reach more than $94\%$, which is comparable to that of a recent related experiment~\cite{Jin}. For example, when determining  whether a state is separable or not, the accuracy of the ML models can reach over $97\%$. When applied to one-way steering and Bell non-locality, the prediction accuracies can be over $96\%$ and $94\%$, respectively. These results are shown in FIG. \ref{distribution}b.

Next, we consider applying the ML models to simultaneously label all four classes of states. The patterns of the states trained through ANN (see SM~\cite{supplementry} for the detailed structure), SVM and DT are shown in FIG. \ref{distribution}c, d and e. The accuracies are $90.11\%$, $90.11\%$ and $85.17\%$, respectively. States in red and blue colors are labeled to be separable and entangled, in green color are labeled to be one-way steerable and in yellow color are labeled to be Bell non-local, respectively. Error bars are estimated from the Poissonian counting statistics, which are smaller than the size of the dots. The inset in FIG. \ref{distribution} is a magnification of the region in the black pane to exhibit error bars.

Gray lines in FIG. \ref{distribution}c-e represent the theoretical boundaries determined by the value of $p$ and $\theta$. Compared with the states labeled by traditional criteria, we can see that the prediction errors of ML models mainly occur at the boundaries, which are marked as ``$\times$". Since the calculation of steering radius is extremely sensitive when $\theta$ is close to 0, $\pi/2$, $\pi$ and $3\pi/2$, we omit these states in the test of the ability of the learning networks to reduce the errors induced by the physical criteria. A more detailed discussion is shown in SM~\cite{supplementry}.

\begin{figure}[t]
	\begin{center}
		\includegraphics[width=1\columnwidth]{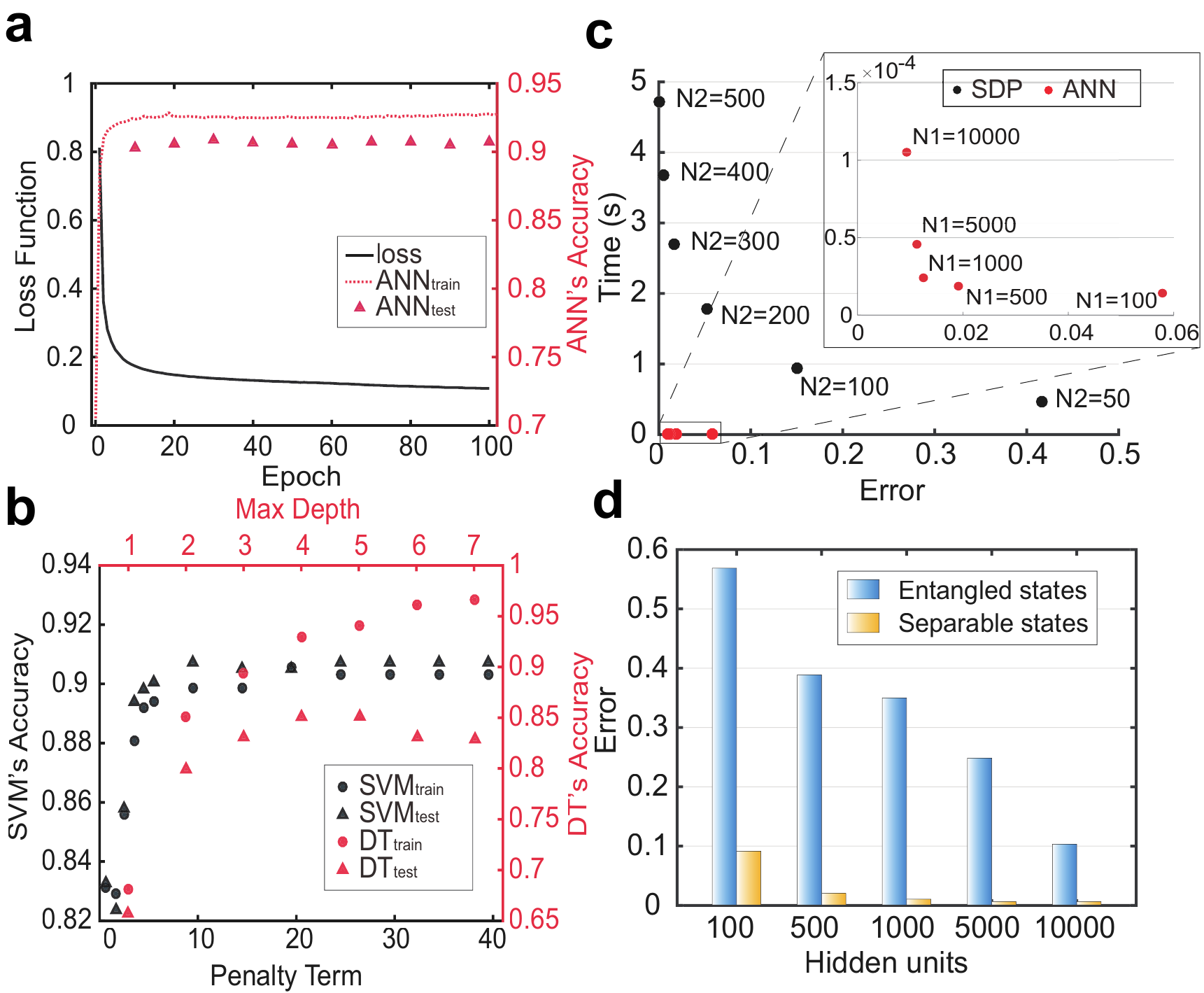}
		\caption{The comparison between three ML methods. \textbf{a}. The performance of ANN versus the Epochs. The black curves represent the value of the loss function and the red dotted line represents the training accuracy. Red triangles show the tested accuracy in different Epochs. \textbf{b}. The performance of DT and SVM versus max tree's depth and error penalty factor with the data marked in black and red colors, respectively. Circles and triangles represent the  training and testing results respectively. \textbf{c.} The relationship among labeling time, error and iterations (N2) of SDP (black); The relationship among labeling time, error and hidden units (N1) of ANN (red). \textbf{d.} The variation of error along with the number of hidden units for distinguishing entangled/separable states.
		}
		\label{curve}
	\end{center}
\end{figure}

\textit{Data analysis}.---The accuracy of the ML methods depends on the settings of several parameters, which are further investigated (see FIG.~\ref{curve}). The variations of the loss function and accuracy as a function of Epochs (stages) for the ANN is shown in FIG. \ref{curve}a. The implementation of ANN is based on the TensorFlow API, where the loss function~\cite{loss} is defined by the categorical cross-entropy. The training process is to minimize the loss function by $RMSprop$~\cite{RMSprop}. The  accuracy is define as the proportion of the number of samples that are correctly labeled by ML models compared with traditional criteria. From the FIG. \ref{curve}a, we found that the loss function decreases rapidly with the training accuracy growing quickly. An epoch is one complete presentation of the data set to be learned to a learning machine. In our work, we set max Epochs to 100. The value of loss function is 0.1477 and the training accuracy is $92.93\%$. The change of Epochs will cause slight disturbance to prediction accuracy. The value of $90.11\%$ is achieved when Epochs=30. 


We further analyzed the impact of penalty term~\cite{penalty} of SVM and DT's maximum tree depth~\cite{tree} on the prediction accuracy, as shown in FIG. \ref{curve}b. In the initial stage, with the increase of tree depth, the prediction ability becomes stronger for DT. When reaching the optimal depth 4, the DT prediction accuracy is $85.17\%$. Larger depth will make DT's structure more complex and cause overfitting. The properties of SVM are similar to that of ANN. When penalty term is greater than 25, the prediction accuracy will converge to $90.11\%$. It's due to the fact that the structure is not sensitive to penalty term in our task. See SM~\cite{supplementry} for the explanations of some jargons.

The run time is an important index to evaluate the performance of the ML algorithms. Given the same computing resources, state tomography requires about 5.28 s, while ANN costs about 0.33 s, and both SVM and DT take a very short time (10$^{-5}$ s) to output the label, indicating that ML models can significantly reduce the time complexity.


\textit{Scalability}.---For systems with a higher dimension, e.g. a pair of qutrits, there is no known efficient way of labeling the training set. In this case, one may rely on numerical means, for example, semi-definite programming (SDP)~\cite{Vandenberghe1996} for labeling. In the following, we shall demonstrate how our ML model can also be applied to learn the labels generated by SDP, using ANN on a pair qutrits. 

In the following, we assume that SDP with 500 iterations for each data point (state) would be sufficient for separating the entangled states for a random set of states; For 100000 pairs of qubits, we found that the accuracy of SDP 500 can reach nearly $100\%$ compared with the PPT criterion. 
Furthermore, we studied the error and run time in classifying 768280 ($90\%$ training + $10\%$ testing) general 2-qutrit states labeled by SDP, which are equally divided between the entangle states and separable states.
The result is shown in FIG. \ref{curve}c (red points), where N1 is the number of hidden units representing the nerual nodes of the hidden layer in the neural network. For comparison, we also plot the run time and error relative to SDP 500 for SDP with different iterations N2 as black points in FIG.~\ref{curve}c. These results show that our ML model can efficiently learn SDP 500 with a small error. 


We also analyze the main error sources of ANN. FIG. \ref{curve}d illustrates the relationship between prediction error and hidden units for different types of 2-qutrit states (8000 entangle states + 12936 separable states). With the increase of hidden units, the entanglement/separable states error decrease. For the same number of hidden units, the ANN prediction error of entangled states is much larger than that of separable states. Therefore, the prediction error of neural network is mainly due to the inaccurate prediction of entangled states.


\textit{Conclusion}.---In this work, we experimentally explored the possibility of the classification problem of multiple quantum correlations by comparing three ML approaches, namely ANN, SVM, and DT. It is shown that, all three methods can be experimentally trained to efficiently learn and classify quantum states, without state tomography. 


Comparing these three ML methods with tomography and traditional criteria, we conclude that the predictive power is ordered as follows: 
by using the traditional criterion as the standard, ANN and SVM has higher accuracies than DT.
In addition, ML models using only partial information of the state, which greatly reduces the number of projection measurements (8 projection measurements for ML features (observables) / 16 projection measurements for tomography in a 2-qubit system). Moreover, predictions of ML models have less time complexity.

Furthermore, we discuss the influence of the data source of the training set~\cite{supplementry}. It suggests 
the necessity of the experimental verification. 


\begin{acknowledgments}
{\textit{Acknowledgement}.---This work was supported by the National Key Research and Development Program of China (Grant No. 2016YFA0302700), the National Natural Science Founda-tion of China (Grants No. 61725504, 11774335 and 11821404), the Key Research Program of Frontier Sciences, Chinese Academy of Sciences (CAS) (Grant No. QYZDY-SSW-SLH003), Science Foundation of the CAS (No. ZDRW-XH-2019-1), Anhui Initiative in Quantum Information Technologies (AHY060300 and AHY020100), the Fundamental Research Funds for the Central Universities (Grant No. WK2030380017 and WK2470000026). 
C.-L. R. was supported by National key research and development program (No. 2017YFA0305200), the Youth Innovation Promotion Association (CAS) (Grant No. 2015317), the National Natural Science Foundation of China (Grant No. 11605205), the Natural Science Foundation of Chongqing (Grants No. cstc2015jcyjA00021 and No. cstc2018jcyjA2509), the Entrepreneurship and Innovation Support Program for Chongqing Overseas Returnees (Grants No. cx2017134 and No. cx2018040).
M.-H. Y. was supported by the National Natural Science Foundation of China (~11875160), the Guangdong Innovative and Entrepreneurial Research Team Program (No.~2016ZT06D348),  Natural Science Foundation of Guangdong Province (2017B030308003), and the Science, Technology and Innovation Commission of Shenzhen Municipality (JCYJ20170412152620376, JCYJ20170817105046702, ZDSYS201703031659262).

M. Y. and C.-L. R. contributed equally to this work.}

\end{acknowledgments}

\end{document}


\renewcommand{\thefigure}{S\arabic{figure}}
\renewcommand\theequation{S\arabic{equation}}

\title{Supplemental Material of `` Experimental Simultaneous Learning of Multiple Non-Classical Correlations"}

\author{Mu Yang}
\affiliation{CAS Key Laboratory of Quantum Information, University of Science and Technology of China, Hefei 230026, People's Republic of China}
\affiliation{CAS Center For Excellence in Quantum Information and Quantum Physics, University of Science and Technology of China, Hefei 230026, People's Republic of China}

\author{Chang-liang Ren}
\affiliation{Center for Nanofabrication and System Integration, Chongqing Institute of Green and Intelligent Technology, 400714, Chinese Academy of Sciences, People’s Republic of China}

\author{Yue-chi Ma}
\affiliation{ Center for Quantum Information, Institute for Interdisciplinary
	Information Sciences, Tsinghua University, Beijing 100084, People's Republic of China}
\affiliation{Shenzhen Institute for Quantum Science and Engineering and Department of Physics, Southern University of Science and Technology, Shenzhen 518055, China}

\author{Ya Xiao}
\affiliation{Department of Physics, Ocean University of China, Qingdao 266100, People's Republic of China}

\author{Xiang-Jun Ye}
\affiliation{CAS Key Laboratory of Quantum Information, University of Science and Technology of China, Hefei 230026, People's Republic of China}
\affiliation{CAS Center For Excellence in Quantum Information and Quantum Physics, University of Science and Technology of China, Hefei 230026, People's Republic of China}

\author{Lu-Lu Song}
\affiliation{Shenzhen Institute for Quantum Science and Engineering and Department of Physics, Southern University of Science and Technology, Shenzhen 518055, China}
\affiliation{Shenzhen Key Laboratory of Quantum Science and Engineering,
	Southern University of Science and Technology, Shenzhen, 518055, China}

\author{Jin-Shi Xu}\email{jsxu@ustc.edu.cn}
\affiliation{CAS Key Laboratory of Quantum Information, University of Science and Technology of China, Hefei 230026, People's Republic of China}
\affiliation{CAS Center For Excellence in Quantum Information and Quantum Physics, University of Science and Technology of China, Hefei 230026, People's Republic of China}

\author{Man-Hong Yung}\email{yung@sustc.edu.cn}
\affiliation{Shenzhen Institute for Quantum Science and Engineering and Department of Physics, Southern University of Science and Technology, Shenzhen 518055, China}
\affiliation{Shenzhen Key Laboratory of Quantum Science and Engineering,
	Southern University of Science and Technology, Shenzhen, 518055, China}
\affiliation{Central Research Institute, Huawei Technologies, Shenzhen, 518129, China}

\author{Chuan-Feng Li}\email{cfli@ustc.edu.cn}
\affiliation{CAS Key Laboratory of Quantum Information, University of Science and Technology of China, Hefei 230026, People's Republic of China}
\affiliation{CAS Center For Excellence in Quantum Information and Quantum Physics, University of Science and Technology of China, Hefei 230026, People's Republic of China}

\author{Guang-Can Guo}
\affiliation{CAS Key Laboratory of Quantum Information, University of Science and Technology of China, Hefei 230026, People's Republic of China}
\affiliation{CAS Center For Excellence in Quantum Information and Quantum Physics, University of Science and Technology of China, Hefei 230026, People's Republic of China}
\maketitle	

In this supplements, we provide and discuss more experimental results. In Sec. (I), we introduce the definition of one-way steering radius in two settings and the structure of artificial neural network (ANN). In Sec. (II), we compare the projection bases of tomography and features measurements. In Sec. (III), we show the performance of predicting experimental states by a well-trained multi-label classifier using simulated states. It efficiently proves the importance of using experimental states as training data but not the simulated state produced by computer. In Sec. (IV), we show how we get the training set and test set. In Sec. (V), we explain some important jargons of machine learning. In Sec. (VI), we introduce Semidefinite programs in detail.

\section{I. One-way steering radius and ANN models}

\subsection{One-way steering radius in two settings}
In the steering task, we can define Bob's conditional states after receiving the measurement results $\kappa|\vec{n}$ from Alice, 
\begin{equation}
\hat{\rho}_{\kappa|\vec{n}}=\dfrac{Tr_{A}[\rho_{AB}(\Pi_{\kappa|\vec{n}}\otimes I)]}{Tr[\rho_{AB}(\Pi_{\kappa|\vec{n}}\otimes I)]}, 
\end{equation}
where $\kappa\in(0,1)$ when Alice measuring along the direction $\vec{n}$, and $\Pi_{\kappa|\vec{n}}=[I+(-1)^{\kappa}\vec{n}\cdot\vec{\sigma}]/2$.  $I$ represents the identity matrix and $\vec{\sigma}=(\sigma_{x},\sigma_{y},\sigma_{z})$ is the Pauli vector.
If Bob's conditional states can be rewritten as a combination of local hidden states ${\rho_{i}}$, 
\begin{equation}
\hat{\rho}_{\kappa|\vec{n}}=\sum_iP(\kappa|\vec{n}, i)p_i\rho_i, 
\end{equation}
there exists a local hidden state model (LHSM) to describe the conditional states, and the steering task fails. The probability distribution $P(\kappa|\vec{n}, i)$ is a stochastic map. 
The steering radius is defined as \cite{sun2016experimental}
\begin{equation}
R_{A\to B}(\rho_{AB})_{\{x,y\}}=min\{max\{L[\rho_{i}]\}\},
\end{equation}
where $L[\rho_{i}]$ denotes the length of Bloch vectors of the states $\rho_{i}$. If $ R_{A\rightarrow B}>1 $, the local hidden states can be located outside of the Bloch sphere, therefor the conditional states obtained on Bob's side can't be described by LHSM. Alice can then steer Bob. The analysis is the same when Bob
wants to steer Alice.

For two-measurement settings, according to the symmetrical property of the steering ellipsoid~\cite{jevtic2014quantum,jevtic2015einstein} of $\rho_{AB}(p,\theta)$, the optimal measurement settings is $\lbrace \vec{x}, \vec{z}\rbrace $ for steering directions. In our work, the state labeled as one-way steerable from Alice to Bob if and only if $ R_{A\rightarrow B}>1 $ and $ R_{B\rightarrow A}\leq 1 $.

\subsection{The structure of ANN model}
ANN is consist of input layer, hidden layer and output layer. The input layer consists of four neurons, each of which correspond to a feature of the input state. We set the input features mentioned as $\vec{x}_{0}$, The intermediate vector $\vec{x}_{1}$  though the hidden layer is generated by the non-linear relation,
\begin{equation}\label{x1}
\vec{x}_{1}={\sigma}_{RL}({W}_{1}\vec{x}_{0}+\vec{\omega}_{1}),
\end{equation}
where ${\sigma}_{RL}$ is the ReLu function in each neuron of the hidden layer, defined as ${\sigma}_{RL}(z)_{i}=max(z_{i},0) (i=1, 2, 3...) $.
The matrix ${W}_{1}$ is the initialized weights and the vector $\vec{\omega}_{1}$  is the bias between the input layer and the hidden layer. Both of them are optimized during the learning process. There are four neurons in the output layer which correspond to four types of quantum correlations, including Bell non-local, one-way steerable, entangeled and separable. The optimal output vector denoted as $\vec{x}_{2}$ is generated though the function,
\begin{equation}\label{x2}
\vec{x}_{2}={\sigma}_{s}({W}_{2}\vec{x}_{1}+\vec{\omega}_{2}),
\end{equation}
where ${\sigma}_{s}$ is the Softmax function in the hidden layer's neurons, defined by ${\sigma}_{s}(z)_{i}=e^{z_{i}}/\sum_{k=1}^{4}e^{z_{k}} (i=1, 2, 3, 4)$. The matrix ${W}_{2}$ is the initialized weights between the hidden layer and the output layer while the vector $\vec{\omega}_{2}$ is the bias. The loss function is categorical cross-entropy, which is written as $-\dfrac{1}{n}\sum_{s}[y_{s}\log a_{s}+(1-y_{s})\log (1-a_{s})]$. 
$s$ means training sample sequence number, $y$ represents the labels defined by the criterion, $a$ means the output labels of ANN and $n$ is the number of training set, respectively. The training process is to minimize the loss function by ANN's optimizer $RMSprop$.

\section{II. Projection bases for tomography and observables measurement}
The number of projection measurements to obtain the input features (partial information) for ML models (8 projection measurements) is half that of quantum state tomography (16 projection measurements) in a 2-qubit system (see Table \ref{bases}). 

\begin{table}[h]
	\centering
	\caption{ Projection bases for tomography and observables measurement}
	\resizebox{\textwidth}{!}{
		\begin{tabular}{cccccccc}	
			\hline\noalign{\smallskip}		
			Tomography & $|HH\rangle\langle HH|$  & $|HV\rangle\langle HV|$  & $|HR\rangle\langle HR|$ & $|HD\rangle\langle HD|$ & $|VD\rangle\langle VD|$ & $|VR\rangle\langle VR|$ & $|VH\rangle\langle VH|$ \\	
			\noalign{\smallskip}
			~ & $|VV\rangle\langle VV|$  & $|RV\rangle\langle RV|$  & $|RH\rangle\langle RH|$ & $|RR\rangle\langle RR|$ & $|RD\rangle\langle RD|$ & $|DD\rangle\langle DD|$ & $|DR\rangle\langle DR|$ \\
			\noalign{\smallskip}
			~ & $|DH\rangle\langle DH|$  & $|DV\rangle\langle DV|$  &  &  &  &  & \\
			\hline\noalign{\smallskip}
			Features (Observables) & $|A_{0}B^{'}_{0}\rangle\langle A_{0}B^{'}_{0}|$ & $|A_{0}^{\perp}B^{'}_{0}\rangle\langle |A_{0}^{\perp}B^{'}_{0}|$ & $|A_{0}B^{'\perp}_{0}\rangle\langle A_{0}B^{'\perp}_{0}|$ & $|A_{0}^{\perp}B^{'\perp}_{0}\rangle\langle A_{0}^{\perp}B^{'\perp}_{0}|$ & $|A_{0}^{' }B_{0}\rangle\langle A_{0}^{'}B_{0}|$ & $|A_{0}^{' \perp}B_{0}\rangle\langle B_{0}^{' \perp}B_{0}|$ & $|A_{0}^{'}B^{\perp}_{0}\rangle\langle A_{0}^{'}B^{\perp}_{0}|$ \\
			\noalign{\smallskip}
			~ & $|A_{0}^{' \perp}B^{\perp}_{0}\rangle\langle A_{0}^{' \perp}B^{\perp}_{0}|$ &  &  &  &  &  & \\
			\noalign{\smallskip}
			\hline
			
		\end{tabular}\label{bases}}	
	\begin{tablenotes}
		\tiny
		\item[1] 
		$|D\rangle= |H\rangle+|V\rangle$; $|R\rangle= |H\rangle-i|V\rangle$; \\
		$|A_{0}\rangle= |H\rangle$;
		$|A_{0}^{'}\rangle=cos\frac{-\pi}{4}|H\rangle+sin\frac{-\pi}{4}|V\rangle$;
		$|B_{0}\rangle= cos\frac{-\pi}{8}|H\rangle+sin\frac{-\pi}{8}|V\rangle$;
		$|B_{0}^{'}\rangle=cos\frac{\pi}{8}|H\rangle+sin\frac{\pi}{8}|V\rangle$;\\
		
	\end{tablenotes}
\end{table}


The observables $\langle{a}_{0}{b}^{\prime}_{0}\rangle$, $\langle{a}^{\prime}_{0}{b}_{0}\rangle$ can be calculated by:
\begin{equation}\label{ab}
\langle{a}_{0}{b}^{\prime}_{0}\rangle=
\dfrac{N_{|A_{0}B^{'}_{0}\rangle\langle A_{0}B^{'}_{0}|}
+N_{|A_{0}^{\perp}B^{'\perp}_{0}\rangle\langle|A_{0}^{\perp}B^{'\perp}_{0}|}
-N_{|A_{0}^{\perp}B^{'}_{0}\rangle\langle A_{0}^{\perp}B^{'}_{0}|}
-N_{|A_{0}B^{'\perp}_{0}\rangle\langle A_{0}^{\perp}B^{'\perp}_{0}|}}
{N_{|A_{0}B^{'}_{0}\rangle\langle A_{0}B^{'}_{0}|}
+N_{|A_{0}^{\perp}B^{'\perp}_{0}\rangle\langle|A_{0}^{\perp}B^{'\perp}_{0}|}
+N_{|A_{0}^{\perp}B^{'}_{0}\rangle\langle A_{0}^{\perp}B^{'}_{0}|}
+N_{|A_{0}B^{'\perp}_{0}\rangle\langle A_{0}^{\perp}B^{'\perp}_{0}|}},
\end{equation}

\begin{equation}\label{ab}
\langle{a}^{\prime}_{0}{b}_{0}\rangle=
\dfrac{N_{|A_{0}^{'}B_{0}\rangle\langle A_{0}^{'}B_{0}|}
+N_{|A_{0}^{' \perp}B^{\perp}_{0}\rangle\langle A_{0}^{' \perp}B^{\perp}_{0}|}
-N_{|A_{0}^{' \perp}B_{0}\rangle\langle B_{0}^{' \perp}B_{0}|}
-N_{|A_{0}^{'}B^{\perp}_{0}\rangle\langle A_{0}^{'}B^{\perp}_{0}|}}
{N_{|A_{0}^{'}B_{0}\rangle\langle A_{0}^{'}B_{0}|}
+N_{|A_{0}^{' \perp}B^{\perp}_{0}\rangle\langle A_{0}^{' \perp}B^{\perp}_{0}|}
+N_{|A_{0}^{' \perp}B_{0}\rangle\langle B_{0}^{' \perp}B_{0}|}
+N_{|A_{0}^{'}B^{\perp}_{0}\rangle\langle A_{0}^{'}B^{\perp}_{0}|}},
\end{equation}

where $N$ represents the photon counts of projection measurements.

\section{III. The performance of predicting experimental states by a well-trained multiple quantum correlation classifier using simulated states}


\begin{figure}[!hbt]
	\begin{center}
		\includegraphics[width=0.8\columnwidth]{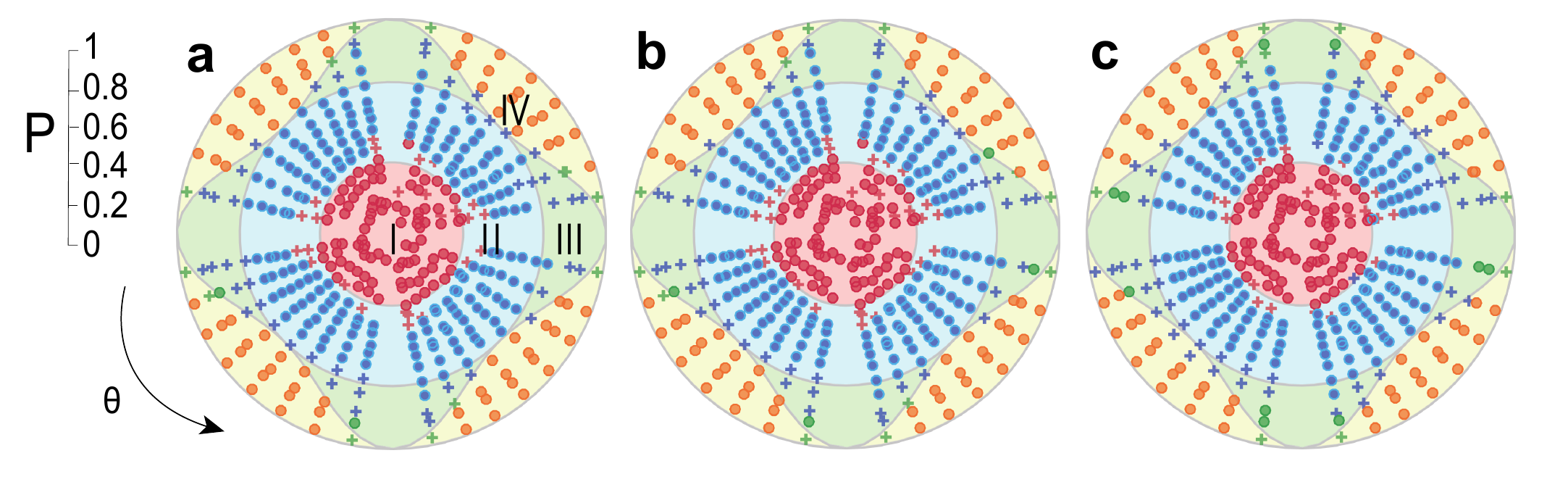}
		\caption{The results of machine-learning models trained by numerical states. States in red and blue colors are determined to be separable and entangled, in green color are determined to be one-way steerable and in yellow color are determined to be Bell nonlocal, respectively. The test results with ANN, SVM and DT are shown in {\bf a.}, {\bf b.} and {\bf c.} with the accuracies are  $80.67\%$, $80.67\%$ and $82.70\%$, respectively. The parameter $\theta$ varies from $0$ to $2\pi$ while the visibility $p$ changes from $0$ to $1$. The space is divided into four parts by the gray theoretical boundary, in which states in I, II, III and IV are theoretically predicted to be separable, entangled, one-way steerable and Bell nonlocal, respectively.}		
		\label{Thory_exp}
	\end{center}
\end{figure}

Here we want to clarify an interesting and important point: can we use a well-trained multiple quantum correlation classifier which is trained by using simulated states to efficiently predict the real experimental data? One may expect that these well-trained multiple quantum correlation classifier can efficiently avoid experimental noise during the training process, and it can also achieve a very high accuracy in the cross validation process. Hence, it should be very effective. But such intuition is incorrect. We generated 1004 different quantum states by computer and constructed three well-trained multiple quantum correlation classifiers with different ML methods. Then taking such quantum correlation classifiers to predict the real experimental data. The accuracies of ANN, SVM and DT are $80.67\%$, $80.67\%$ and $82.70\%$ , which are shown in FIG. \ref{Thory_exp}a-c, respectively. The prediction error on the boundary is very large, and non of a one-way steerable state is predicted. The theoretically trained ML models are failures for experimental states. It illustrates that the sources of the training data and the test data heavily affect the efficiency of the classifiers. This also explain the necessity of our experimental demonstration.

\section{IV. Training and test set}

\begin{figure}[!hbt]
	\begin{center}
		\includegraphics[width=0.9\columnwidth]{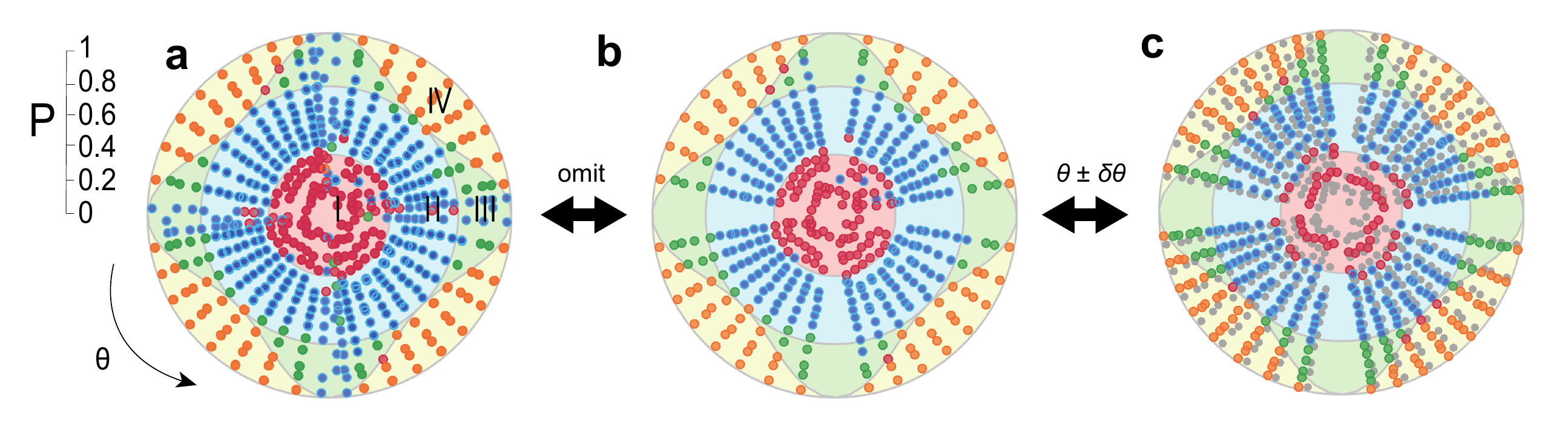}
		\caption{The results including states with $\theta$ close to $0$, $\pi/2$, $\pi$, $3\pi/2$. The parameter $\theta$ varies form $0$ to $2\pi$ while the visibility $p$ changes from $0$ to $1$. The space is divided into 4 parts by the gray boundary, in which I, II, III and IV represent the separable state, entangled state,  one-way steerable state, and Bell nonlocal state, respectively. {\bf a.} The quantum states are labeled by different quantum correlation criteria with tomography data. States in red and blue colors are determined to be separable and entangled, in green color are determined to be one-way steerable and in yellow color are determined to be Bell nonlocal, respectively. {\bf b.} The state distribution when the states of $\theta$ closing to 0, $\pi/2$, $\pi$, $3\pi/2$ are omitted. {\bf c.} The training set is different from the test set. The gray points represent the test set.}
		\label{local_whole}
	\end{center}
\end{figure}

When the parameter $\theta$ is close to $0$, $\pi/2$, $\pi$, $3\pi/2$, the steering ellipsoid will become very small. So the criterion of steering radius will be extremely unstable. The result is shown in FIG. \ref{local_whole}a, in which the one-way steerable states (green color) determined from ($p$, $\theta$) are found not to be one-way steerable (blue color) determined from the reconstructed density matrixes. Therefore, we do not take into the consideration of the states when $\theta$ close to 0, $\pi/2$, $\pi$, $3\pi/2$, which is shown in FIG. \ref{local_whole}b. In our work, the training set ($p$, $\theta$) is different with the test set ($p$, $\theta$+$\delta\theta$), which is shown in FIG. \ref{local_whole}c, where the gray points represent test set.


\section{V. Explanations for the jargons of machine learning}

Here we add some  explanations for the jargons of machine learning, including Epoch, loss function, penalty term, maximum tree depth and overfitting:\\

\textbf{Epoch}: An epoch is one step in training a neural network. Concretely, when a neural network is trained on every training samples each time, we say that one epoch is finished. Training process usually consists more than one epoch.\\

\textbf{Loss function}: A loss function is a function that maps an event onto a real number intuitively representing some ``cost” associated with the event. An optimization problem seeks to minimize the loss function.\\

\textbf{Penalty term}: Penalty term (C parameter) is one of the important parameters in support-vector machine (SVM), which tells the SVM optimization how much you want to avoid misclassifying each training example. An appropriate C parameter can improve SVM performance.\\

\textbf{Maximum tree depth}: Maximum tree depth means the length of the longest path from the tree root to a leaf of the decision tree (DT). Limiting the maximum tree depth can improve the performance of DT.\\

\textbf{Overfitting}: Overfitting is the production of an analysis that corresponds too closely or exactly to a particular set of data, and may therefore fail to fit additional data or predict future observations reliably.\\

\section{VI. Semidefinite programs for distinguishing separable and entangled states}

\subsection{Background}
One of the most popular criterion for ``witnessing" entangled states is based on taking partial transposition, introduced by Peres~\cite{Peres1996}. For any density matrix $\rho$, with the matrix elements defined by $\rho_{ik,jl} = \langle i | \otimes \langle k | \rho | j \rangle \otimes | l \rangle $, its partial transpose is defined as $ \rho_{ik, jl}^{T_A} = \rho_{jk,il}$. Thus if the states $\rho$ is separable, it must have a positive partial transpose. Furthermore, any state for which $\rho^{T_A}$ is not positive semidefinite is necessarily entangled. Horodecki et al proved that positive partial transpose (PPT) is both necessary and sufficient for identifying separability for a pair of qubits $\mathscr{H}_2 \otimes \mathscr{H}_2 $ and qubit-qutrit systems~$\mathscr{H}_2 \otimes \mathscr{H}_3$~\cite{Horodecki1996}. However, for higher dimensional systems, there exist entangled states that cannot be identified by the PPT criterion~\cite{Horodecki1997}.

Alternatively, due to the approach known as semidefinite program (SDP)~\cite{Vandenberghe1996} for convex optimization problem, one may theoretically and numerically~\cite{Doherty2002} construct a separability criteria based on the properties of separable states.

A separable state $\rho$ can be written as a convex combination of product states $\rho = \sum p_i | \psi_i \rangle \langle \psi_i | \otimes | \phi_i \rangle \langle \phi_i |$, where $| \psi_i \rangle $ and $ | \phi_i \rangle $ are on the spaces $\mathscr{H}_A$ and $ \mathscr{H}_B $. Let $ \tilde{\rho}$ defined on $\mathscr{H}_A \otimes \mathscr{H}_B \otimes \mathscr{H}_A $, is given by
\begin{equation}
\tilde{\rho} = \sum p_i | \psi_i \rangle \langle \psi_i | \otimes | \phi_i \rangle \langle \phi_i | \otimes | \psi_i \rangle \langle \psi_i |.
\end{equation}
Thus, the state $\tilde{\rho}$ satisfies following properties. First, the state $\tilde{\rho}$ is an extension of $\rho$ (the partial trace over the third party is equal to $\rho$). Second, the state $\tilde{\rho}$ is symmetric under interchanging the two copies of $\mathscr{H}_A$. Finally, the extension $\tilde{\rho}$ is a tripartite separable state.

Thus, if a state $\rho$ on $\mathscr{H}_A \otimes \mathscr{H}_B $ is separable, there exists an extension $\tilde{\rho} $ on $ \mathscr{H}_A \otimes \mathscr{H}_B \otimes \mathscr{H}_A $ such that $\tilde{\rho}^{T_A} \geq 0 $ and $\tilde{\rho}^{T_B} \geq 0 $. We can also generalize the criterion to an arbitrary copies of  $ \mathscr{H}_A $ and $ \mathscr{H}_B $. That is, if a state $\rho$ on $\mathscr{H}_A \otimes \mathscr{H}_B $ is separable, there is an extension $\tilde{\rho}= \sum p_i | \psi_i \rangle \langle \psi_i |^{\otimes k} \otimes | \phi_i \rangle \langle \phi_i |^{\otimes l}$ on the symmetric subspace $ \mathscr{H}_A^k \otimes \mathscr{H}_B^l  $ such that $ \tilde{\rho}$ has a positive partial transpose for all partitions.

Thus, given an state $\rho$, the usual PPT criterion is first used to test. If the test fails, the state is entangled; otherwise, the state can be entangled or separable. For the latter case, we look for an extension $\tilde{\rho}$ of $\rho$ to three parties such that it satisfies all possible partial transpose, which can be solved through SDP introduced in the next. If extension $\tilde{\rho}$ can not be found, we say that the state is entangled; otherwise, we can look for an entension for four parties.

\subsection{Semidefinite programs for separability problem}
The above problem can been solved through SDP, which is basically expressed as
\begin{equation}\label{SDP}
\begin{split}
minimize &  \quad c^{T} \textbf{x}, \\
subject \quad to & \quad F(\textbf{x}) \geq 0,
\end{split}
\end{equation}
where $c$ is a given vector, $\textbf{x} = (x_1,...,x_m)$ and $F(\textbf{x}) = F_0 + \sum_i F_i x_i$, for some fixed $n$-by-$n$ Hermitian matrices $F_j$. The inequality in the second line of Eq. \ref{SDP} means that the matrix $ F(\textbf{x}) $ is semidefinite. For a particular case when $c=0$, the problem reduces to whether it is possible to find $ \textbf{x} $ such that $F(\textbf{x}) $ is semidefinite. This turns into a feasibility problem and lightens a separability criteria.

We only consider the problem of searching for an extension of $\rho$ to three parties here. Let $\{  \sigma^{A}_i \}_{ i= 1,...,d^2_{A}}$ and $\{  \sigma^{B}_j \}_{ j= 1,...,d^2_{B}}$ be bases for the space of Hermitian matrices that operate on $ \mathscr{H}_A $ and $ \mathscr{H}_B $ respectively, such that they satisfy
\begin{equation}
Tr(\sigma^{X}_i \sigma^{X}_j) = \alpha \delta_{ij} \quad and \quad Tr(\sigma^{X}_i) = \alpha \delta_{i1},
\end{equation}
where $X$ stands for $A$ or $B$, and $\alpha$ is some constant. Thus we can write $\rho = \sum_{ij} \rho_{ij} \sigma^{A}_i \sigma^{B}_j $ ,with $\rho_{ij} = \alpha^{-2} Tr [\rho \sigma^{A}_i \sigma^{B}_j]$. The extension $\tilde{\rho}$ can be written in a similar way
\begin{equation}
\tilde{\rho} = \sum \limits_{ij,i<k} \tilde{\rho}_{kji} \{ \sigma^{A}_i \otimes \sigma^{B}_j \otimes \sigma^{A}_k + \sigma^{A}_k \otimes \sigma^{B}_j \otimes \sigma^{A}_i \} + \sum\limits_{kj} \{ \sigma^{A}_k \otimes \sigma^{B}_j \otimes \sigma^{A}_k \}.
\end{equation}
We also need to satisfy that the trace of $\tilde{\rho}$ for the third party is $ \rho$, that is $Tr_C [\tilde{\rho} ] = \rho$, which leads to $\tilde{\rho}_{ij1} = \rho_{ij}$. The remaining components of $\tilde{\rho}$ will be variables in our SDP. Thus we require that the extension $\tilde{\rho}$ and its partial transpose must be semidefinite. For example, if we want the condition $\tilde{\rho} \geq 0$ to be transformed into $F( \textbf{x}) = F_0 + \sum_i F_i x_i \geq 0$, we can define
\begin{equation}
\begin{split}
F_0 & = \sum_{j} \rho_{1j} \sigma^{A}_1 \otimes \sigma^{B}_j \otimes \sigma^{A}_1 + \sum_{i=2,j=1} \rho_{ij} \{ \sigma^{A}_i \otimes \sigma^{B}_j \otimes \sigma^{A}_1 + \sigma^{A}_1 \otimes \sigma^{B}_j \otimes \sigma^{A}_i \}, \\
F_{iji} & = \sigma^{A}_i \otimes \sigma^{B}_j \otimes \sigma^{A}_i, \quad i\geq 2, \\
F_{ijk} & = ( \sigma^{A}_i \otimes \sigma^{B}_j \otimes \sigma^{A}_k + \sigma^{A}_k \otimes \sigma^{B}_j \otimes \sigma^{A}_i), \quad k > i \geq 2. \\
\end{split}
\end{equation}
The coefficients $\tilde{\rho}_{ijk} ( k \neq 1, k\geq i)$ play the role of variable $\textbf{x}$. Moreover, positivity of the partial transpose leads to two more inequalities, $\tilde{\rho}^{T_A} \geq 0 $ and $\tilde{\rho}^{T_B} \geq 0 $.

Therefore, the whole problem, whether a state $\rho$ is separable or not, turns into finding $ \tilde{\rho}_{ijk} (k \neq 1, k\geq i)$ with $ G= \tilde{\rho} \otimes \tilde{\rho}^{T_A} \otimes \tilde{\rho}^{T_B} \geq 0 $.  If such $ \tilde{\rho}_{ijk} (k \neq 1, k\geq i)$  don't exist, the state $\rho$ is entangled; If we cannot find such $ \tilde{\rho}_{ijk} (k \neq 1, k\geq i)$, the state $\rho$ may be separable or entangled.

Numerical SDP solvers have been introduced in detail in~\cite{Vandenberghe1996}.  Given the symmetric subspace $ \mathscr{H}_A^k \otimes \mathscr{H}_B^l  $, the iteration number for checking this criteria is no more than $O (d^{13k/2}_A  d^{13l/2}_B )$. Specifically, in our work, we apply SDP in the space $ \mathscr{H}_A \otimes \mathscr{H}_B \otimes  \mathscr{H}_A $.

{}